\documentstyle[11pt]{article}

\begin{document}

\newtheorem{theorem}{Theorem}[section]
\newtheorem{lemma}[theorem]{Lemma}
\newtheorem{defin}[theorem]{Definition}
\newtheorem{rem}[theorem]{Remark}
\newtheorem{cor}[theorem]{Corollary}

 %
 %
 % MMM MMM MMM MMM MMM
 %
 %
\def\m{\mu} \def\l{\lambda} \def\e{\epsilon} \def\r{\rho}

\def\be{\begin{equation}}
\def\ee{\end{equation}}
\def\bea{\begin{eqnarray}}
\def\eea{\end{eqnarray}}
\def\beas{\begin{eqnarray*}}
\def\eeas{\end{eqnarray*}}

\def\dt{\partial_t}
\def\dx{\partial_x}
\def\dv{ \partial_v }

\renewcommand\div{\mbox{\rm div}\;}

\def\fn{ \baselineskip = 0pt
\vbox{\hbox{\hspace*{3pt}\tiny $\circ$}\hbox{$f$}} \baselineskip = 12pt\!}

\def\N{{\rm I\kern-.1567em N}}
\def\No{\N_0}
\def\I{{\rm I\kern-.1567em I}}
\def\R{{\rm I\kern-.1567em R}}

\def\supp{\mbox{\rm supp}}
\def\vol{\mbox{\rm vol}}
\def\n#1{\vert #1 \vert}
\def\nn#1{\Vert #1 \Vert}

\def\prf{\noindent
         {\em Proof :\ }}
\def\prfe{\hspace*{\fill} $\Box$

\smallskip \noindent}

\title{ A regularity theorem for solutions of the spherically symmetric
Vlasov-Einstein system}

\author{Gerhard Rein\thanks
        {Mathematisches Institut der Universit\"at M\"unchen,
        Theresienstr.\ 39, W8000 M\"unchen 2, Germany.},
        Alan D.~Rendall\thanks
        {Max-Planck-Institut f\"ur Astrophysik,
        Karl-Schwarzschild-Str.\ 1, W-8046 Garching bei M\"unchen.},
        and Jack Schaeffer\thanks
        {Department of Mathematics, Carnegie-Mellon University,
        Pittsburgh, PA 15213, USA\newline
        Research supported in part by NSF DMS 9101517}}
\date{}
\maketitle
\begin{abstract}
We show that if a solution of the spherically
symmetric Vlasov-Einstein system
develops a singularity at all then the
first singularity has to appear at the center of symmetry.
The main tool is an estimate which shows that a solution is global
if all the matter remains away from the
center of symmetry.
\end{abstract}
 %
 %
 % INTRO INTRO INTRO
 %
 %
\section{Introduction}
This paper is concerned with the long-time behaviour of solutions of the
spherically symmetric Vlasov-Einstein system. In [4] a continuation
criterion was obtained for solutions of this system, and it was shown
that for small initial data the corresponding solution exists globally
in time. In the following we investigate what happens for large
initial data. In the coordinates used in [4] the equations to be
solved are as follows:
\be \label{v}
\dt f+ e^{\m-\l}\frac{v}{\sqrt{1+v^2}} \cdot \dx f -
\Bigl(\frac{x\cdot v}{r} \dot\l + e^{\m-\l}
\sqrt{1+v^2}\m' \Bigr) \frac{x}{r} \cdot \dv f=0,
\ee
\be \label{e1}
e^{-2\l}(2r\l'-1)+1=8\pi r^2\r,
\ee
\be \label{e2}
e^{-2\l}(2r\m'+1)-1=8\pi r^2 p,
\ee
\be \label{rdef}
\r(t,x)=\int\sqrt{1+v^2} f(t,x,v)\,dv,
\ee
\be \label{pdef}
p(t,x)=\int \left(\frac{x\cdot v}{r}\right)^2
f(t,x,v)\frac{dv}{\sqrt{1+v^2}}.
\ee
Here $x$ and $v$ belong to $\R^3$, $r:=|x|$, $x\cdot v$ denotes the
usual inner product of vectors in $\R^3$, and $v^2:=v\cdot v$.
The distribution function
$f$ is assumed to be invariant under simultaneous rotations of $x$
and $v$, and hence $\r$ and $p$ can be regarded as functions of $t$
and $r$. Spherically symmetric functions of $t$ and $x$ will be
identified with functions of $t$ and $r$ whenever it is convenient.
In particular $\l$ and $\m$ are regarded as functions of $t$
and $r$, and the dot and prime denote derivatives with respect to $t$
and $r$ respectively. It is assumed that $f(t)$ has compact support
for each fixed $t$. We are interested in regular asymptotically flat
solutions which leads to the boundary conditions that
\be \label{bc}
\l(t,0)=0,\ \ \ \lim_{r\to\infty}\m(t,r)=0,
\ee
for each fixed $t$.

The main aim of this paper is to show that if singularities ever
develop in solutions of the system (\ref{v})--(\ref{pdef})
with the above boundary conditions
then the first singularity must be at the center of symmetry. In
order to do this we consider solutions of (\ref{v})--(\ref{pdef})
on a certain
kind of exterior region with different boundary conditions and
prove that for the modified problem there exists a global in time
solution for any initial data. One of the most interesting
implications of the results of \cite{RR1} is that the naked singularities
in spherically symmetric solutions of the Einstein equations
coupled to dust can be cured by passing to a slightly different
matter model, namely that described by the Vlasov equation,
in the case of small initial data. The results of this paper
strengthen this conclusion to say that shell-crossing singularities
are completely eliminated.

The following remarks put our results in context. For the
Vlasov-Poisson system, which is the non-relativistic analogue
of the Vlasov-Einstein system, it is known that global existence
holds for boundary conditions which
are the analogue of the requirement of asymptotic flatness in
the relativistic case \cite{LP,P,S} and also in a cosmological setting
\cite{RR2}. No symmetry assumptions are
necessary. For the relativistic Vlasov-Poisson system with
an attractive force spherically symmetric solutions with negative
energy develop singularities in finite time \cite{GS}. It is easy
to show that in these solutions the first singularity
occurs at the center of symmetry. On the other hand it was also
shown in \cite{GS} that spherically symmetric solutions of the relativistic
Vlasov-Poisson system with a repulsive force never develop
singularities. The latter are in one-to-one correspondence with
spherically symmetric solutions of the Vlasov-Maxwell system.

The paper is organized as follows.
Sect.~2 contains the main estimates together with a proof that
they imply global existence in the case that all the matter
remains away from the center. In Sect.~3 a local existence
theorem and continuation criterion for the exterior problem
are proved. It is then shown that the estimates of Sect.~2
imply a global existence theorem for the exterior problem.
Finally, in Sect.~4, these results are combined to give the
main theorem.

\section{The restricted regularity theorem}

\setcounter{equation}{0}

The goal of this section is to show that a solution may be extended as long
as $f$ vanishes in a neighborhood of the center of symmetry. Consider an
initial datum $\fn \geq 0$ which is spherically symmetric, $C^1$,
compactly supported, and satisfies
\be \label{incon}
\int_{|x|<r}\int \sqrt{1+v^2}\fn(x,v)\, dv\, dx < r/2
\ee
for all $r>0$. By Theorem 3.1 of \cite{RR1}
a regular solution $(f,\l ,\m )$
of the system (\ref{v})--(\ref{pdef}) with boundary conditions
(\ref{bc}) and initial datum $\fn$ exists on $[0,T[\times \R^6$
for some $T>0$.

\begin{theorem} \label{glex}
Let $\fn$ and $T$ be as above ($\,T$ finite). Assume there
exists $\e >0$ such that
\be \label{van}
f(t,x,v)=0\ \mbox{if}\ 0\leq t<T \ \mbox{and}\ |x|\leq \e .
\ee
Then $(f,\l ,\m )$ extends to a regular solution on $[0,T'[$
for some $T'>T.$
\end{theorem}

Define
\be \label{p}
P(t) := \sup \Bigl\{|v|:(x,v)\in \supp f(t)\Bigr\},
\ee
then Theorem 3.2 of \cite{RR1} states that Theorem \ref{glex} above follows
once $P(t)$ is shown to be bounded on $[0,T[$.
We need a few other facts from \cite{RR1}:
\be \label{lambda}
e^{-2\l }=1-2m/r
\ee
where
\be \label{mass}
m(t,r) := 4\pi \int_0^rs^2\r (t,s)ds
\ee
(equations (2.11) and (2.12) of \cite{RR1}). Also
\be \label{dlambda}
\dot \l =-4\pi \;re^{\m + \l}j
\ee
where
\be \label{curr}
j(t,r) := \int \frac{x\cdot v}{r} f(t,x,v) \,dv
\ee
(equations (3.37) and (3.38) of \cite{RR1}).

The following notation will be used:
\[
u:=|v|,\ w:=r^{-1}x\cdot v,\ F := |x\wedge v|^2=r^2u^2-(x\cdot v)^2 .
\]
Differentiation along a characteristic
of the Vlasov equation is denoted by $D_t,$ so
\[
D_tx=e^{\m -\l }\frac{v}{\sqrt{1+u^2}}
\]
and
\[
D_tv=-\left(\frac{x\cdot v}{r} \dot \l + e^{\m -\l }\sqrt{1+u^2}\,
\m'\right)\frac{x}{r}.
\]
It follows that
\be \label{fcon}
D_t F=0 ,
\ee
\be \label{dr}
D_tr=e^{\m -\l }\frac{w}{\sqrt{1+u^2}},
\ee
and
\[
D_t w=-r^{-2}(D_tr)x\cdot v+r^{-1}(D_tx)\cdot v+r^{-1}x\cdot (D_tv).
\]
Substitution for $D_tr,D_tx,$ and $D_tv$ and simplification yields
\be \label{dw1}
D_t w  = \frac{F}{r^3 \sqrt{1+u^2}} e^{\m -\,\l }
- w \dot \l  - e^{\m -\l }\sqrt{1+u^2}\, \m '.
\ee
The letter $C$ will denote a generic constant which changes from line to
line, and may depend only on $\fn, \e$, and $T$.

Now we make a few preliminary estimates. The values of $f$ are conserved
along characteristics so
\[
0\leq f\leq \sup \,\fn\,=C.
\]
Also we claim that
\be \label{mcons}
\int \r (t,x)\,dx=\int \r (0,x)\,dx=C
\ee
To show this multiply (\ref{v}) by $\sqrt{1+v^2}$ and integrate
in $v,$ which yields (after simplification)
\beas
0 & = & \dt \r +e^{\m -\l } \div \Bigl( \int f v \,dv\Bigr)
 +(\r +p)\dot \l + 2 j e^{\m -\l }\m' \\
& = & \dt\r +\div \Bigl(e^{\m -\l }\int fv\,dv \Bigr)
+(\r +p)\dot \l +j e^{\m -\l }(\m'+\l').
\eeas
Now substituting (\ref{dlambda}) and (\ref{e1}), (\ref{e2}) this becomes
\be \label{conlaw}
0=\dt\r +\div \Bigl(e^{\m -\l }\int f v\,dv \Bigr)
\ee
and (\ref{mcons}) follows. Also by (\ref{mass})
\be \label{mb}
0\leq m(t,r)\leq \int \r (t,x)\,dx \leq C,\ r\geq 0.
\ee

It follows from (\ref{e1}), (\ref{e2}) that
\[
\m'+\l'\geq 0
\]
and from (\ref{bc}) and (\ref{lambda}) that
\[
\lim_{r \to \infty}(\m +\l )=0,
\]
so
\[
\m -\l \leq \m +\l \leq 0
\]
and
\be \label{beml}
e^{\m -\l }\leq e^{\m +\l }\leq 1.
\ee

Note that
\[
0\leq F\leq C
\]
on the support of $f$, so
\be \label{u}
u^2=w^2+\frac{F}{r^2} \leq w^2+\frac{C}{\e^2}=w^2+C.
\ee
Hence we will focus on $w$. Define
\be \label{pi}
P_i(t)  :=  \inf \Bigl\{w:\exists x,v \ \mbox{with}\
f(t,x,v)\neq 0 \ \mbox{and}\ w=r^{-1}x\cdot v\Bigr \}
\ee
and
\be \label{ps}
P_s(t)  :=  \sup \Bigl\{w:\exists x,v \ \mbox{with}\
f(t,x,v)\neq 0\ \mbox{and}\ w=r^{-1}x\cdot v \Bigr\},
\ee
then if $P_i(t)$ and $P_s(t)$ are bounded, it follows that $P(t)$ is
bounded. Also note that
\be \label{meas}
\mbox{measure}\; \Bigl\{v:(x,v)\in \supp f(t) \Bigr\}
\leq \pi \, C\, \e^{-2} \Bigl( P_s(t)-P_i(t) \Bigr),\ \n{x} \geq \e.
\ee

Next we focus on the characteristic equation for $w$. Note first that by
(\ref{e2}) and (\ref{lambda})
\beas
\m' & = & \frac{1}{2r}\Bigl(e^{2\l }(8\pi r^2 p+1)-1\Bigr)
 = \frac{1}{2r} e^{2\l }\Bigl(8\pi r^2p+1-1+
\frac{2m}{r}\Bigr) \\
& = & e^{2\l }(r^{-2}m+4\pi rp).
\eeas
Using this and (\ref{dlambda}) in (\ref{dw1}) yields
\be \label{dw2}
D_tw  = \frac{F}{r^3\sqrt{1+u^2}} e^{\m -\l }
-  r^{-2}\sqrt{1+u^2}\, e^{\m +\l }m
+  4\pi re^{\m +\l }(w j-\sqrt{1+u^2}\, p)
\ee
By (\ref{van}) and (\ref{beml}) we may bound the first term of (\ref{dw2}) by
\[
0  \leq \frac{F}{r^3\sqrt{1+u^2}} e^{\m -\l }
\leq  \e ^{-3}C=C
\]
on the support of $f$. By (\ref{van}), (\ref{u}), (\ref{beml}), and
(\ref{mb}) we may bound the
second term of (\ref{dw2}) by
\[
0 \leq  r^{-2}\sqrt{1+u^2}\, e^{\m +\l }m
\leq  \e ^{-2}\sqrt{C+w^2} C\leq C\sqrt{C+w^2}
\]
on the support of $f$. Hence, (\ref{dw2}) becomes
\bea
-C\sqrt{C+w^2}   +  4\pi re^{\m +\l }(wj-\sqrt{1+u^2}\, p)
& \leq & D_tw \label{dw3} \\
& \leq & C+4\pi re^{\m +\l }(wj-\sqrt{1+u^2}\, p). \nonumber
\eea
On the support of $f$
\[
0\leq 4\pi re^{\m +\l}\leq C
\]
so we must consider the quantity $w j-\sqrt{1+u^2}\, p$. Let us denote
\[
\tilde w := r^{-1}x\cdot \tilde v,\ \tilde u := |\tilde v|,\
\tilde F := |x\wedge \tilde v|^2=r^2\tilde u^2-(x\cdot \tilde v)^2.
\]
Then
\be \label{wj}
w j-\sqrt{1+u^2}\, p
= \int f(t,x,\tilde v) \Bigl(w\tilde w-
\sqrt{1+u^2} \frac{\tilde w^2}{\sqrt{1+\tilde u^2}}\Bigr)\, d\tilde v.
\ee

The next step is to use (\ref{dw3}) and (\ref{wj})
to derive an a upper bound for $w$
(on the support of $f$), and hence for $P_s(t)$. This may be done without a
bound on $P_i(t).$ Then (\ref{dw3}), (\ref{wj}), and the bound for
$P_s(t)$ will be
used to derive a lower bound for $w$, and hence for $P_i(t).$ Then by
(\ref{u}) and (\ref{p}) a bound for $P(t)$ follows,
and the solution may be extended.

To bound $P_s(t)$ suppose
\[
P_s(t)>0
\]
and consider $w$ (in $\supp f$) with
\[
w>0.
\]
For $\tilde w\leq 0$ we have
\[
w\tilde w-\sqrt{1+u^2} \frac{\tilde w^2}{\sqrt{1+\tilde u^2}}\leq 0.
\]
For $\tilde w>0$ we have
\beas
w \tilde w - \sqrt{1+u^2} \frac{\tilde w^2}{\sqrt{1+\tilde u^2}}
&=&
\frac{\tilde w}{\sqrt{1+\tilde u^2}}
\frac{w^2 (1+ \tilde u^2)-\tilde w^2 (1+u^2)}
{w \sqrt{1 + \tilde u^2} + \tilde w \sqrt{1+u^2} } \\
&=&
\frac{\tilde w}{\sqrt{1+\tilde u^2}}
\frac{w^2 (1+\tilde F r^{-2}) - \tilde w^2 (1+F r^{-2})}
{w \sqrt{1+\tilde u^2} + \tilde w \sqrt{1+u^2}}.
\eeas
Note that in the last step a term of ``$w^2\tilde w^2$'' canceled, which is
crucial. Hence
\[
w\tilde w- \sqrt{1+u^2} \frac{\tilde w^2}{\sqrt{1+\tilde u^2}}
\leq
\frac{\tilde w}{\sqrt{1+\tilde u^2}}
\frac{w^2 (1+C \e^{-2})}{w \sqrt{1+\tilde u^2}}
\leq
C \frac{w \tilde w}{1+\tilde w^2}.
\]
Now using the above and (\ref{meas}) we have
\beas
\int f\Bigl( w\tilde w-\sqrt{1+u^2}
\frac{\tilde w^2}{\sqrt{1+\tilde u^2}} \Bigr) d\tilde v
&\leq &
\int_{0<\tilde w<P_s(t)} f\;C\;w\frac{\tilde w}{1+\tilde w^2} d\tilde v \\
&\leq&
\pi\,C\,\e ^{-2} \int_0^{P_s(t)} C\;w\frac{\tilde w}{1+\tilde w^2}
d\tilde w \\
&=&
C\;w\;\ln (1+P_s^2(t))\leq C\;P_s(t)\ln (1+P_s^2(t)),
\eeas
and by (\ref{wj}) and (\ref{dw3})
\be \label{dw4}
D_tw\leq C+C\;P_s(t)\ln (1+P_s^2(t)),
\ee
for $w>0$ in the support of $f$. Denote the values of $w$ along a
characteristic by $w(\tau )$ and let
\[
t_0 := \inf \Bigl\{\tau \geq 0:w(s)\geq 0\ \mbox{for}\ s\in ]\tau ,t[\Bigr\}.
\]
Then either $t_0=0$ or $w(t_0)=0$ and in either case
\[
w(t_0)\leq C.
\]
Hence by (\ref{dw4})
\beas
w(t)
&\leq&
w(t_0)+\int_{t_0}^t\Bigl(C+CP_s(\tau )\ln (1+P_s^2(\tau ))\Bigr)d\tau \\
&\leq&
C+C\;\int_{t_0}^t P_s(\tau )\ln (1+P_s^2(\tau ))d\tau .
\eeas
Defining
\[
\overline{P}_s(\tau ) := \max \{0,P_s(\tau )\}
\]
we may write
\[
w(t)\leq C+C\;\int_0^t\overline{P}_s(\tau )\ln (1+\overline{P}_s^2(\tau))d\tau
\]
and hence
\[
\overline{P}_s(t) \leq
C+C\;\int_0^t\overline{P}_s(\tau )\ln (1+\overline{P}_s^2(\tau ))d\tau .
\]
We assumed that $P_s(t)>0,$ but note that the last inequality is valid in
all cases. It now follows that
\be
\label{psest}
P_s(t)\leq \overline{P}_s(t)\leq \exp (e^{ct})\leq C
\ee
on $t\in [0,T[.$

To bound $P_i(t)$ from below suppose
\[
P_i(t)<0
\]
and consider $w$ (in support $f$) with
\[
P_i(t)<w<0.
\]
For $\tilde w\leq 0$ we have
\beas
w \tilde w - \sqrt{1+u^2} \frac{\tilde w^2}{ \sqrt{1+\tilde u^2} }
&= &
\frac{\tilde w}{ \sqrt{1+\tilde u^2} }
\frac{w^2 (1+\tilde Fr^{-2}) -\tilde w^2 (1+F r^{-2})}
{w \sqrt{1+\tilde u^2} + \tilde w \sqrt{1+u^2} } \\
&= &
\frac{|\tilde w|}{ \sqrt{1+\tilde u^2} }
\frac{w^2 (1+\tilde F r^{-2}) - \tilde w^2 (1+Fr^{-2})}
{|w| \sqrt{1+\tilde u^2} + |\tilde w| \sqrt{1+u^2}} \\
&\geq &
\frac{|\tilde w|}{\sqrt{1+\tilde u^2}}
\frac{(-\tilde w^2) (1+Fr^{-2})}
{|\tilde w| \sqrt{1+u^2}} \\
&\geq&
\frac{-\tilde w^2\,(1+C\e ^{-2})}
{\sqrt{1+\tilde w^2} \sqrt{1+w^2}}\\
&\geq&
 -C\frac{|\tilde w|}{\sqrt{1+w^2}}.
\eeas
For $w<0<\tilde w\leq P_s(t)$ we have (using (\ref{psest}))
\beas
w\tilde w-\sqrt{1+u^2}\frac{\tilde w^2}{\sqrt{1+\tilde u^2}}
&\geq&
P_s(t)w - \sqrt{1+w^2+F r^{-2}}\frac{\tilde w^2}{\sqrt{1+\tilde w^2}}\\
&\geq&
C w - |\tilde w| \sqrt{1+C\e ^{-2}+w^2} \\
&\geq&
C w - P_s(t) \sqrt{C+w^2} \\
&\geq&
 -C \sqrt{C+w^2}.
\eeas
Hence (using (\ref{meas}) as before)
\beas
&&\int f\Bigl(w\tilde w -
\sqrt{1+u^2}\frac{\tilde w^2}{\sqrt{1+\tilde u^2}} \Bigr) d\tilde v \\
&&\qquad \geq
\int_{\tilde w\leq 0} f \Bigl( -C \frac{|\tilde w|}{\sqrt{1+w^2}}\Bigr)
d\tilde v +
\int_{\tilde w>0} f\Bigl(-C \sqrt{C+w^2}\Bigr)d\tilde v \\
&&\qquad
\geq
-\frac{C}{\sqrt{1+w^2}} \pi\, C\,\e ^{-2}
\int_{-P_i(t)}^0|\tilde w| d\tilde w
-C \sqrt{C+w^2} \pi\, C\, \e ^{-2}
\int_0^{\overline{P}_s(t)}d\tilde w \\
&& \qquad
=
- C P_i^2(t)\frac{1}{\sqrt{1+w^2}} - C\overline{P}_s(t)\sqrt{C+w^2} .
\eeas
Now by (\ref{wj}), (\ref{dw3}), and (\ref{psest})
\beas
D_tw
&\geq&
- C \sqrt{C+w^2}
- C P_i^2(t)\frac{1}{\sqrt{1+w^2}}
- C \overline{P}_s(t) \sqrt{C+w^2}  \\
&\geq&
- C\sqrt{C+w^2} - C P_i^2(t) \frac{1}{\sqrt{1+w^2}}.
\eeas
Since we have assumed $0>w>P_i(t),$ it is convenient to write this as
\beas
D_t(w^2) & = & 2wD_tw \\
&\leq&
C (-w) \sqrt{C+w^2} +C P_i^2(t)\frac{(-w)}{\sqrt{1+w^2}} \\
&\leq&
C |P_i(t)| \sqrt{C+P_i^2(t)} + C P_i^2(t) \\
&\leq&
C + C P_i^2(t).
\eeas
As before define
\[
t_1 := \inf \Bigl\{\tau \geq 0:w(s)\leq 0\ \mbox{for}\ s \in ]\tau ,t[\Bigr\}
\]
then
\[
0\geq w(t_1)\geq -C
\]
so
\beas
w^2(t) & \leq & C+\int_{t_1}^t(C+CP_i^2(\tau ))d\tau \\
& \leq & C+C\int_0^tP_i^2(\tau )d\tau .
\eeas
It follows that
\be \label{piest}
P_i^2(t)\leq C+C\int_0^tP_i^2(\tau )d\tau
\ee
if $P_i(t)<0.$ But if $P_i(t)\geq 0$ then
\[
0\leq P_i(t)\leq P_s(t)\leq C
\]
so (\ref{piest}) holds in this case, too.
Now by Gronwall's inequality it follows that
\[
P_i^2(t)\leq e^{ct}\leq C
\]
on $[0,T[.$ Finally a bound for $P(t)$ follows from (\ref{u}) and
(\ref{p}), and the proof is complete.
 %
 % SECT 3
 %

\section{The exterior problem}
\setcounter{equation}{0}

If $r_1$ and $T$ are positive real numbers, define the exterior
region
\[
W(T,r_1) := \{(t,r):0 \leq t<T, r \geq r_1+t\}.
\]
In this section
the initial value problem for (\ref{v})--(\ref{pdef})
will be studied on a region
of this kind. Consider an initial datum $\fn(x,v)$ defined on the
region $|x|\geq r_1$ which is non-negative, compactly supported,
$C^1$, and spherically symmetric. The first of the boundary
conditions (\ref{bc}) cannot be used in the case of an exterior region,
and so it will be replaced as follows. Let $m_\infty$ be any number
greater than $4\pi\int_{r_1}^\infty r^2\r(0,r)dr$. For any
solution of (\ref{v})--(\ref{pdef}) on $W(T,r_1)$ define
\be \label{nmdef}
m(t,r) := m_\infty-4\pi\int_r^\infty s^2\r(t,s)ds.
\ee
Provided $m_\infty$ satisfies the above inequality, the quantity
$m(0,r)$ is everywhere positive. The replacement for (\ref{bc}) is:
\be \label{nbc}
e^{-2\l(t,r)}=1-2m(t,r)/r,\ \ \ \lim_{r\to\infty}\m(t,r)=0.
\ee
The first of these conditions is a combination of (\ref{e1})
with a choice of boundary condition. Note that if a
solution of the original problem with boundary conditions (\ref{bc})
is restricted to $W(T,r_1)$ then it will satisfy (\ref{nbc}) provided
$m_\infty$ is chosen to be equal to the ADM mass of the solution
on the full space. Just as in the local existence theorem in [4],
a further restriction must be imposed on the initial datum. In
this case it reads
\be \label{restr1}
m_\infty-\int_{|x|\geq r}\int \sqrt{1+v^2} \fn(x,v)\,dv\,dx < r/2,\ r \geq r_1.
\ee
This is of course necessary if (\ref{nbc}) is to hold on the initial
hypersurface. The nature of the solutions to be constructed is
encoded in the following definition.

\smallskip
\noindent
{\bf Definition} A solution $(f,\l,\m)$ of (\ref{v})--(\ref{pdef})
on a region of the form $W(T,r_1)$ is called {\em regular} if
\begin{itemize}
\item[(i)]
$f$ is non-negative, spherically symmetric, and $C^1$, and $f(t)$ has
compact support for each $t\in [0,T[$
\item[(ii)]
$\l\geq0$, and $\l$, $\m$, $\l'$, and
$\m'$ are $C^1$.
\end{itemize}

\smallskip
\noindent
A local existence theorem can now be stated.

\begin{theorem} \label{locex}
Let $m_\infty>0$ be a fixed real number.
Let $\fn\geq0$ be a spherically symmetric function on the region
$|x|\geq {r_1}$ which is $C^1$ and has compact support. Suppose that
(\ref{restr1}) holds for all $r\geq r_1$ and that
\be \label{restr2}
\int_{|x|\geq r_1}\int \sqrt{1+v^2} \fn(x,v)\,dv\,dx<m_\infty.
\ee
Then there exists a unique regular spherically symmetric solution of
(\ref{v})--(\ref{pdef})
on a region $W(T,r_1)$ with $f(0) = \fn$ and satisfying (\ref{nbc}).
\end{theorem}

\prf
This is similar in outline to the proof of Theorem 3.1 of
[4] and thus will only be sketched, with the differences compared to
that proof being treated in more detail. Define $\l_0(t,r)$ and
$\m_0(t,r)$ to be zero. If $\l_n$ and $\m_n$ are defined on
the region $W(T_n,r_1)$ then $f_n$ is defined to be the solution
of the Vlasov equation with $\l$ and $\m$ replaced by
$\l_n$ and $\m_n$ respectively and initial datum $\fn$. In
order that this solution be uniquely defined it is necessary to
know that no characteristic can enter a region of the form $W(T,r_1)$
except through the initial hypersurface which is guaranteed if
$\l_n\geq 0$ and $\m_n\leq 0$ on the region of interest. This
will be proved by induction. If $f_n$ is given then $\l_{n+1}$ and
$\m_{n+1}$ are defined to be the solutions of the field equations
with $\r_n$ and $p_n$ constructed from $f_n$ rather than $f$. The
quantities $\l_{n+1}$ and $\m_{n+1}$ are defined on the maximal
region $W(T_{n+1},r_1)$ where
$0 < m_n(t,r) < r/2$ so that $\l_{n+1}$ can be defined by the first
equation in (\ref{nbc}) and is positive on $W(T_{n+1},r_1)$; note that
$0 < m_n(t,r) < r/2$ for $r$ large and $0 < m_n(0,r) < r/2,\ r\geq r_1$
by assumption on $\fn$ so that $T_{n+1} > 0$ by continuity.
It can be shown straightforwardly that the iterates
$(\l_n,\m_n,f_n)$ are well-defined and regular for all $n$.
The most significant difficulty in proving the corresponding statement in
[4] was checking the differentiability of various quantities at the
center of symmetry, and in the exterior problem the center of symmetry
is excluded. The next step is to show that there exists some $T>0$
so that $T_n\geq T$ for all $n$ and that the quantities $\l_n$,
$\dot\l_n$, and $\m_n'$ are uniformly bounded in $n$ on the
region $W(T,r_1)$. Let $L_n(t,r) := r^{-1}(1-e^{-2\l_n(t,r)})^{-1}$.
For $t\in [0,T_n[$ define
\beas
P_n(t)&:=&\sup\{|v|:(x,v)\in\supp f_n(t)\},           \\
Q_n(t)&:=&\|e^{2\l_n(t)}\|_\infty + \|L_n(t)\|_\infty.
\eeas
Now it is possible to carry out the same sequence of estimates as in [4]
to get a differential inequality for $\tilde P_n(t) := \max_{0\leq k\leq n}
P_k(t)$ and $\tilde Q_n(t) := \max_{0\leq k\leq n} Q_k(t)$ which is
independent of $n$, and this gives the desired result. The one point
which is significantly different from what was done in [4] is the
estimate for $\dot\l_n$. There a partial integration in $r$
must be carried out, and in the exterior problem the limits of
integration are changed; nevertheless the basic idea goes through.
The remainder of the proof is almost identical to the proof of
Theorem 3.1 of [4]. It is possible to bound $\m_n'$
and $\dot\l_n'$ on the region $W(T,r_1)$ and then show that
the sequence $(\l_n,\m_n,f_n)$ converges uniformly to a regular
solution of (\ref{v})--(\ref{pdef})
on that region. Moreover this solution is unique.
\prfe

Next a continuation criterion will be derived. By the maximal
interval of existence for the exterior problem is meant the largest
region $W(T,r_1)$ on which a solution exists with given initial
data and the parameters $r_1$ and $m_\infty$ fixed.

\begin{theorem} \label{conti}
Let $(f,\l,\m)$ be a regular solution of the
reduced system (\ref{v})--(\ref{pdef})
on $W(T,r_1)$ with compactly supported initial
datum $\fn$. If $T<\infty$ and $W(T,r_1)$ is the maximal interval
of existence then $P$ is unbounded.
\end{theorem}

\prf
Note first that, just as in the case of the problem in
the whole space, the reduced equations (\ref{v})--(\ref{pdef}) imply that the
additional field equation (\ref{dlambda}) is satisfied.
The conservation law (\ref{conlaw}) can be rewritten in the form
\be \label{conlaw2}
\dt (r^2\r)+\partial_r(e^{\m-\l}r^2j)=0.
\ee
Integrating (\ref{conlaw2}) in space shows that the minimum value of
$m$ at time $t$ (which occurs at the inner boundary of $W(T,r_1)$)
is not less than at $t=0$. Hence $L$ is bounded on the whole region.
Suppose now that $P$ is bounded. From
(\ref{dlambda}) it follows that $\dot\l$ is bounded, and if $T$ is finite
this gives an upper bound for $\l$. Combining this with the
lower bound for $\l$ already obtained shows that $Q$ is bounded.
This means that the quantities which influence the size of the
interval of existence are all bounded on $W(T,r_1)$, and it follows
that the solution is extendible.
\prfe

Combining Theorem \ref{conti}
with the estimates of Sect.\ 2 gives a proof that
the solution of the exterior problem corresponding to an initial
datum of the kind assumed in Theorem \ref{locex} exists globally in time
i.\ e.\ that $T$ can be chosen to be infinity in the conclusion of
Theorem \ref{locex}. For we know that it suffices to bound $P$, and the
estimates bound the momentum $v$ along any characteristic on
which $r$ is bounded below. Moreover the inequality $r\geq r_1$
holds on $W(\infty,r_1)$.

\section{The regularity theorem}
\setcounter{equation}{0}

This section is concerned with the initial value problem for
(\ref{v})--(\ref{pdef}) on
the whole space with boundary conditions (\ref{bc}).

\begin{theorem} \label{reg}
Let $(f,\l,\m)$ be a regular solution of the reduced system
(\ref{v})--(\ref{pdef}) on a time interval $[0,T[$.
Suppose that there exists an open neighborhood $U$
of the point $(T,0)$ such that
\be \label{vbound}
\sup\{|v|:(t,x,v)\in\supp f\cap (U\times\R^3)\}<\infty.
\ee
Then $(f,\l,\m)$ extends to a regular solution on $[0,T'[$
for some $T'>T$.
\end{theorem}

\prf Suppose that the condition (\ref{vbound}) is satisfied. Since the
equations are invariant under time translations it can be assumed
without loss of generality that $T$ is as small as desired. Choosing
$T$ sufficiently small ensures that $U$ contains all points with
$(T-t)^2+r^2<4T^2$ and $0\leq t< T$. Now let $r_1<T$. Then
$[0,T[\times\R^3\subset U\cup W(T,r_1)$.
Let $\fn$ be the restriction of $f$ to the hypersurface $t=0$.
Restricting $\fn$ to the region $|x|\geq r_1$ gives an initial datum for
the exterior problem on $W(T,r_1)$. Let $m_\infty$ be the ADM mass of
$\fn$. There are now two cases to be considered, according to whether
$f$ vanishes in a neighborhood of the point $(T,0)$ or not.
If it does then by doing a time translation if necessary it can be
arranged that the matter stays away from the center on the whole
interval $[0,T[$ so that the results of Sect.\ 2 are applicable.
It can  be concluded that the solution extends to a larger time
interval in this case. If $f$ does not vanish in a neighborhood of
$(T,0)$ then by doing a time translation it can be arranged that
(\ref{restr2}) holds.
In this case  the results of the previous section show that
there exists a global solution on $W(\infty,r_1)$ satisfying (\ref{nbc}).
The extended solution must agree on $W(T,r_1)$ with the solution we
started with. Thus a finite upper bound is obtained for $|v|$ on the
part of the support of $f$ over $W(T,r_1)$. But by assumption (\ref{vbound})
we already have a bound of this type on the remainder of the support
of $f$. Hence
\[
\sup\{|v|:(t,x,v)\in\supp f\}<\infty.
\]
Applying Theorem 3.2 of [4] now shows that the solution is extendible
to a larger time interval in this case too.
\prfe

There is another way of looking at this result. Suppose that
$(f,\l,\m)$ is a regular solution of (\ref{v})--(\ref{pdef})
on the whole space
and that $[0,T[$ is its maximal interval of existence. Then by
a similar argument to the above the solution extends in a $C^1$
manner to the set $([0,T]\times\R^3)\setminus\{(T,0)\}$. Thus
if a solution of (\ref{v})--(\ref{pdef})
develops a singularity at all the first
singularity must be at the center.


\begin{thebibliography}{10}

\bibitem{GS}
Glassey, R.\ T., Schaeffer, J.: On symmetric solutions of the
relativistic Vlasov-Poisson system. Commun.\ Math.\ Phys.\ {\bf 101},
459--473 (1985)

\bibitem{LP}
Lions, P.-L., Perthame, B.: Propagation of moments and regularity
for the three dimensional Vlasov-Poisson system. Invent.\ Math.\
{\bf 105}, 415--430 (1991)

\bibitem{P}
Pfaffelmoser, K.: Global classical solutions of the Vlasov-Poisson
system in three dimensions for general initial data. J.\ Diff.\ Eq.\
{\bf 95}, 281--303 (1992)

\bibitem{RR1}
Rein, G., Rendall, A.\ D.: Global existence of solutions of the
spherically symmetric Vlasov-Einstein system with small initial
data. Commun.\ Math.\ Phys.\ {\bf 150}, 561--583 (1992)

\bibitem{RR2}
Rein, G., Rendall, A.\ D.:
Global existence of classical solutions to the Vlasov-Poisson system
in a three dimensional, cosmological setting.
Preprint 1993

\bibitem{S}
Schaeffer, J.: Global existence of smooth solutions of the
Vlasov-Poisson system in three dimensions. Commun.\ Partial Diff.\
Eq.\ {\bf 16} 1313--1336 (1991)

\end{thebibliography}
\end{document}